\documentclass[preprintnumbers,amsmath,amssymb,nofootinbib,superscriptaddress,floatfix,showkeys,twocolumn]{revtex4-1}
\usepackage[utf8]{inputenc}
\usepackage{graphicx,color}
\usepackage{amsmath,amssymb}
\usepackage{hyperref}
\usepackage{epstopdf}
\usepackage{slashed}
\usepackage{soul}
\usepackage{bm}
\usepackage{amssymb}
\usepackage[normalem]{ulem}
\usepackage[caption=false]{subfig}
\usepackage{changes}

\newcommand{\lsim}{\mathrel{\mathop{\kern 0pt \rlap
  {\raise.2ex\hbox{$<$}}}
  \lower.9ex\hbox{\kern-.190em $\sim$}}}
\newcommand{\gsim}{\mathrel{\mathop{\kern 0pt \rlap
  {\raise.2ex\hbox{$>$}}}
  \lower.9ex\hbox{\kern-.190em $\sim$}}}

\newcommand{\mev}{\ensuremath{\,\mathrm{MeV}}}

\DeclareUnicodeCharacter{2212}{\ensuremath{-}}
\definechangesauthor[name={Cui}, color=blue]{Cui}
\definechangesauthor[name={yyli}, color=purple]{yyli}

\begin{document}
\preprint{APS/123-QED}

\title{Primordial magnetic field as a common solution of nanohertz gravitational waves
and the Hubble tension}
\author{Yao-Yu Li}
\affiliation{Key Laboratory of Dark Matter and Space Astronomy,
Purple Mountain Observatory, Chinese Academy of Sciences, Nanjing 210023, China}
\affiliation{School of Astronomy and Space Science, University of Science and Technology of China, Hefei 230026, China}
\author{Chi Zhang}
\affiliation{Key Laboratory of Dark Matter and Space Astronomy,
Purple Mountain Observatory, Chinese Academy of Sciences, Nanjing 210023, China}
\affiliation{School of Astronomy and Space Science, University of Science and Technology of China, Hefei 230026, China}
\author{Ziwei Wang}
\affiliation{Key Laboratory of Dark Matter and Space Astronomy,
Purple Mountain Observatory, Chinese Academy of Sciences, Nanjing 210023, China}
\author{Ming-Yang Cui}
\affiliation{Key Laboratory of Dark Matter and Space Astronomy,
Purple Mountain Observatory, Chinese Academy of Sciences, Nanjing 210023, China}

\author{Yue-Lin Sming Tsai}
\email{smingtsai@pmo.ac.cn}
\affiliation{Key Laboratory of Dark Matter and Space Astronomy,
Purple Mountain Observatory, Chinese Academy of Sciences, Nanjing 210023, China}
\affiliation{School of Astronomy and Space Science, University of Science and Technology of China, Hefei 230026, China}
\author{Qiang Yuan}
\email{yuanq@pmo.ac.cn}
\affiliation{Key Laboratory of Dark Matter and Space Astronomy,
Purple Mountain Observatory, Chinese Academy of Sciences, Nanjing 210023, China}
\affiliation{School of Astronomy and Space Science, University of Science and Technology of China, Hefei 230026, China}
\author{Yi-Zhong Fan}
\affiliation{Key Laboratory of Dark Matter and Space Astronomy,
Purple Mountain Observatory, Chinese Academy of Sciences, Nanjing 210023, China}
\affiliation{School of Astronomy and Space Science, University of Science and Technology of China, Hefei 230026, China}

\begin{abstract}
The origin of interstellar and intergalactic magnetic fields remains largely unknown. 
One possibility is that they are related to the primordial magnetic fields (PMFs) produced by, for instance, the phase transitions of the early Universe. 
In this paper, the PMF-induced turbulence generated at around the QCD phase transition epoch—the characteristic magnetic field strength $B_{\rm ch}^* \sim \mathcal{O}(1)~\rm{\mu G}$ and coherent length scale $\ell_{\rm ch}^* \sim \mathcal{O}(1)~\rm{pc}$—can naturally accommodate nanohertz gravitational waves reported by pulsar timing array (PTA) collaborations.
Moreover,
the evolution of the PMFs to the recombination era with the form of $B_{\rm ch}\sim \ell_{\rm ch}^{-\alpha}$ 
can induce baryon density inhomogeneities, alter the recombination history, and alleviate the tension of the Hubble parameter $H_0$ and the matter
clumpiness parameter $S_8$ between early- and late-time measurements for $0.88\leq \alpha \leq 1.17$ (approximate 95\% credible region based on three PTA likelihoods). This allowed range of $\alpha$ is for the first time obtained by data-driven approach. The further evolved PMFs may account for the $\sim {\cal O}(10^{-16})$ Gauss extragalactic magnetic field inferred with GRB 221009A.
\end{abstract}

\maketitle
\section{Introduction}
A signal of stochastic gravitational wave background (SGWB) with frequencies around nanohertz (nHz) 
is a powerful probe of several astrophysical and physical problems~\cite{1978SvA....22...36S,Detweiler:1979wn,Lasky:2015lej}. 
In recent years, several pulsar timing arrays (PTAs) reported the positive detection of candidate 
power-law signals in the data~\cite{NANOGrav:2020bcs,Goncharov:2021oub,Chen:2021rqp,Antoniadis:2022pcn}.
Very recently, the analyses of the Hellings-Downs (HD) correlation~\cite{Hellings:1983fr} of the 
timing residuals give evidence in support of the SGWB nature of the power-law excess
\cite{NANOGrav:2023gor,Antoniadis:2023ott,Reardon:2023gzh,Xu:2023wog}.
The significance of the HD correlation obtained from NANOGrav, EPTA, PPTA, and CPTA is approximately
$3\sigma$, $3\sigma$, $2\sigma$, and $4.6\sigma$, respectively.
The fitted power law parameters, i.e., the amplitude and spectral index, are
$A_{\rm GWB}=6.4^{+4.2}_{-2.7}\times10^{-15}$ and $\gamma=3.2^{+0.6}_{-0.6}$ for NANOGrav,
$\log_{10} A_{\rm GWB}=-14.54^{+0.28}_{-0.41}$ and $\gamma=4.19^{+0.73}_{-0.63}$ for EPTA,
$A_{\rm GWB}=3.1^{+1.3}_{-0.9}\times10^{-15}$ and $\alpha = -0.45^{+0.20}_{-0.20}$ ($\alpha = \frac{3-\gamma}{2}$) for PPTA,
and $\log_{10} A_{\rm GWB}=-14.4^{+1.0}_{-2.8}$ with $\gamma<6.6$ for CPTA.
These results represent the breakthrough opening a new window for observing the Universe
with gravitational waves (GWs).

Besides the astrophysical origin of the SGWB from the orbital motions of supermassive binary
black holes~\cite{Rajagopal:1994zj,Jaffe:2002rt,Wyithe:2002ep,Sesana:2004sp}, it is of great
interest in possible connection with many new physics processes in the early Universe, such as
inflation~\cite{Grishchuk:2005qe,Vagnozzi:2020gtf,Benetti:2021uea,Vagnozzi:2023lwo,Ashoorioon:2022raz}, phase transitions
\cite{1987MNRAS.229..357D,Kosowsky:2001xp,Brandenburg:2021tmp,Neronov:2020qrl,Caprini:2010xv,Kosowsky:2001xp,Kahniashvili:2009mf,Mazumdar:2018dfl,Addazi:2020zcj,NANOGrav:2021flc,Xue:2021gyq,Kobakhidze:2017mru,Athron:2023xlk},
cosmic strings~\cite{Blasi:2020mfx,Ellis:2020ena,Lazarides:2021uxv, 
Buchmuller:2020lbh,Chang:2021afa,Bian:2022tju,EPTA:2023hof,Samanta:2020cdk,Datta:2020bht},
domain walls~\cite{Liu:2020mru,Bian:2022qbh}, or primordial black 
holes~\cite{Vaskonen:2020lbd,DeLuca:2020agl,Kohri:2020qqd,Domenech:2020ers,Papanikolaou:2023cku,Chen:2019xse}.
It has been proposed that magnetohydrodynamic (MHD) turbulence generated by the primordial magnetic fields (PMFs) can also produce the SGWB~\cite{Caprini:2001nb, Caprini:2006jb,RoperPol:2022iel,Auclair:2022jod}. 
If the PMFs were initially produced via, e.g., the QCD phase transition, the induced SGWB 
would fall within the detectable range of PTAs. The corresponding comoving length is about 
1~pc, inversely proportional to the temperature around $100~\mev$. 
As the Universe evolves to the recombination epoch, the PMFs can 
induce baryon density fluctuations. These baryon density inhomogeneities alter the 
standard cosmological evolution by affecting the recombination history.
Consequently, it may affect the Hubble constant $H_0$ and the matter clumpiness parameter
$S_8$ inferred from the cosmic microwave background (CMB) data~\cite{Planck:2018vyg}.
As a result, it may solve or alleviate \cite{Jedamzik:2020krr,Jedamzik:2018itu} the tension from 
late-time measurements (e.g., \cite{Riess:2021jrx,Reid:2019tiq,Wong:2019kwg}). 

In light of the first detection of SGWB from the PTAs, we study the PMF scenario
with the purpose to account for the nHz SGWB and to alleviate the $H_0$ and
$S_8$ tension simultaneously.~\footnote{See also Ref.~\cite{Bian:2022qbh} for a domain wall network which produces the nanohertz SGWB and largely alleviates the Hubble tension.} We assume an {evolutionary relation} of 
$B_{\rm ch}\sim \ell_{\rm ch}^{-\alpha}$, where $B_{\rm ch}$ is the comoving characteristic magnetic field strength and $\ell_{\rm ch}$ is the comoving scale, 
{to bridge the early time
when PMFs were produced and the recombination epoch.}
The SGWB data are used to constrain the initial parameters of the PMFs. 
They are then implemented as a likelihood to perform a Bayesian global analysis, 
together with the cosmological likelihoods including the Planck CMB anisotropies~\cite{Planck:2018vyg}, 
the baryon acoustic oscillation (BAO) \cite{BOSS:2016wmc,Beutler:2011hx,Ross:2014qpa},
and the local measurements of the Hubble constant (hereafter H3) \cite{Riess:2021jrx,Reid:2019tiq,Wong:2019kwg}. As will be shown in detail later,
we find that the PMFs model can solve these two important problems simultaneously. If this is the case, we also, for the first time, obtain the evolution properties
of the PMFs at different epochs of the early Universe. 

%%%%%%%%%%%%%%%%%%%%%%%%%%%%%%%%%%%%%%%%%%%%%%%%%%%%%%%%%%%%%%%%%%%%%%%5
\section{SGWB from PMFs}
% {\it SGWB from PMFs.}
%%%%%%%%%%%%%%%%%%%%%%%%%%%%%%%%%%%%%%%%%%%%%%%%%%%%%%%%%%%%%%%%%%%%%%%5
\label{PTA_constrain_PMFs}
The PMF is treated as a Gaussian random field with a spectrum described by the same method as~\cite{RoperPol:2022iel,Auclair:2022jod}. 
The spectrum of the GWs produced from the initial time $\tau_*$ to the end time $\tau_{\rm end}$~\footnote{The physical quantity with a superscript or subscript `$\ast$', `end', and `0' represents the value at the initial time $\tau_*$, the end time $\tau_{\rm end}$, and present time, respectively.}  
at the scale $k$ can be written as~\cite{RoperPol:2022iel}
\begin{align}
  &\Omega_{\rm GW}(k, \tau_\text{end}) \simeq 3 \left( \frac{k}{k_{\rm ch}^*}\right) \Omega_{\rm M}^*\frac{\mathcal{C}(\alpha)}{\mathcal{A}^2(\alpha)}p_{\Pi}\left( \frac{k}{k_{\rm ch}^*}\right) \nonumber\\
  &  \times\bigg\{
    \begin{array}{lr}
        \ln^2(1 + \mathcal H_*\delta \tau_\text{end}), &\text{if } \rm{k}\delta\tau_{\text{end}} < 1,\\
        \ln^2(1 + \mathcal{H_*}/k), & \text{if } \rm{k}\delta\tau_{\text{end}} \ge 1,
    \end{array}
\end{align}
where $\Omega_{\rm M}^*=\frac{1}{2}(B_{\rm ch}^*)^2$ is the total normalized magnetic energy density and $k_{\rm ch}^*=2\pi/l_{\rm ch}^*$. The comoving Hubble frequency $\mathcal{H_*} \simeq 1.12\times10^{-8}(T_*/100~\rm{MeV})$, where we set $T_*=100 \mev$ and $g_*=10$ referring to the temperature and the relativistic degrees of freedom during the QCD phase transition epoch. 
Two constants $\mathcal{C}(\alpha)$ and $\mathcal{A}(\alpha)$ have been calculated in Ref.~\cite{RoperPol:2022iel}. 
Also, $\delta\tau_{\rm end}=\tau_{\rm end}-\tau_*$ is the duration of the GW source.  
From the MHD simulation~\cite{RoperPol:2022iel}, 
we have $\delta\tau_{\rm end} = 0.184\mathcal{H_*}^{-1}+1.937\delta \tau_{\rm e}$, 
where $\delta \tau_{\rm e} = (\sqrt{1.5\Omega_M^*}k_{ch}^*)^{-1}$ is the eddy turnover time. The parameter $p_{\Pi}\left( k/k_{\rm ch}^*\right)$ is defined as $P_{\Pi}^*(k)/P_{\Pi}^*(0)$, where $P_{\Pi}^*(k)$ is the anisotropic stress power spectral density.

After $\tau_{\rm end}$, the sources stop acting and the GWs propagate through the Universe freely 
while the energy density decreases due to cosmic expansion. 
Therefore, we have the GW today~\cite{RoperPol:2022iel}
\begin{equation}
\begin{gathered}
 h^{2}\Omega_{\mathrm{GW}}^0(k)=\left(\frac{a_{\mathrm{end}}}{a_0}\right)^4\left(\frac{H_{\mathrm{end}}}{H_0}\right)^2 h^{2} \Omega_{\mathrm{GW}}\left(k, \tau_{\mathrm{end}}\right) \\
\quad \simeq 3.5 \times 10^{-5}~\Omega_{\mathrm{GW}}\left(k, \tau_{\mathrm{end}}\right)\left(\frac{10}{g_{\mathrm{end}}}\right)^{\frac{1}{3}}
,
\label{th_omggw}
\end{gathered}
\end{equation}
where $a_{\rm end}$ and $a_{0}$ correspond to the scale factors. The Hubble constant $H_{\rm end}$ has been calculated in Ref.~\cite{Kolb:1990vq}, and $H_0=100 h~\rm{km\ s^{-1}~Mpc^{-1}}$ with $h\simeq 0.68$~\cite{Planck:2018vyg}. 
The entropic degrees of freedom $g_{\rm end}$ equals $g_*$.
The SGWB spectrum, denoted as $h^2\Omega_{\rm GW}^0(f)$ with frequency $f$, can be linked to the timing
residuals $\rho(f)$ that are directly measured by PTA experiments\cite{NANOGRAV:2018hou}, 
\begin{align}
     & \rho(f)=\frac{1}{4\pi^2f_{\rm yr}}\left( \frac{f}{f_{\rm yr}}\right)  ^{-3/2}h_{\rm c}(f), \\
     & h_{\rm c}(f)=\frac{H_0}{\pi f}\sqrt{\frac{3}{2}h^2\Omega_{\rm GW}^0(f)},
\end{align}
where $f_{\rm yr} \simeq 3.17 \times 10^{-8} \ {\rm Hz}$.

\begin{figure*}[!htb]
    \centering
    \includegraphics[scale = 0.45]{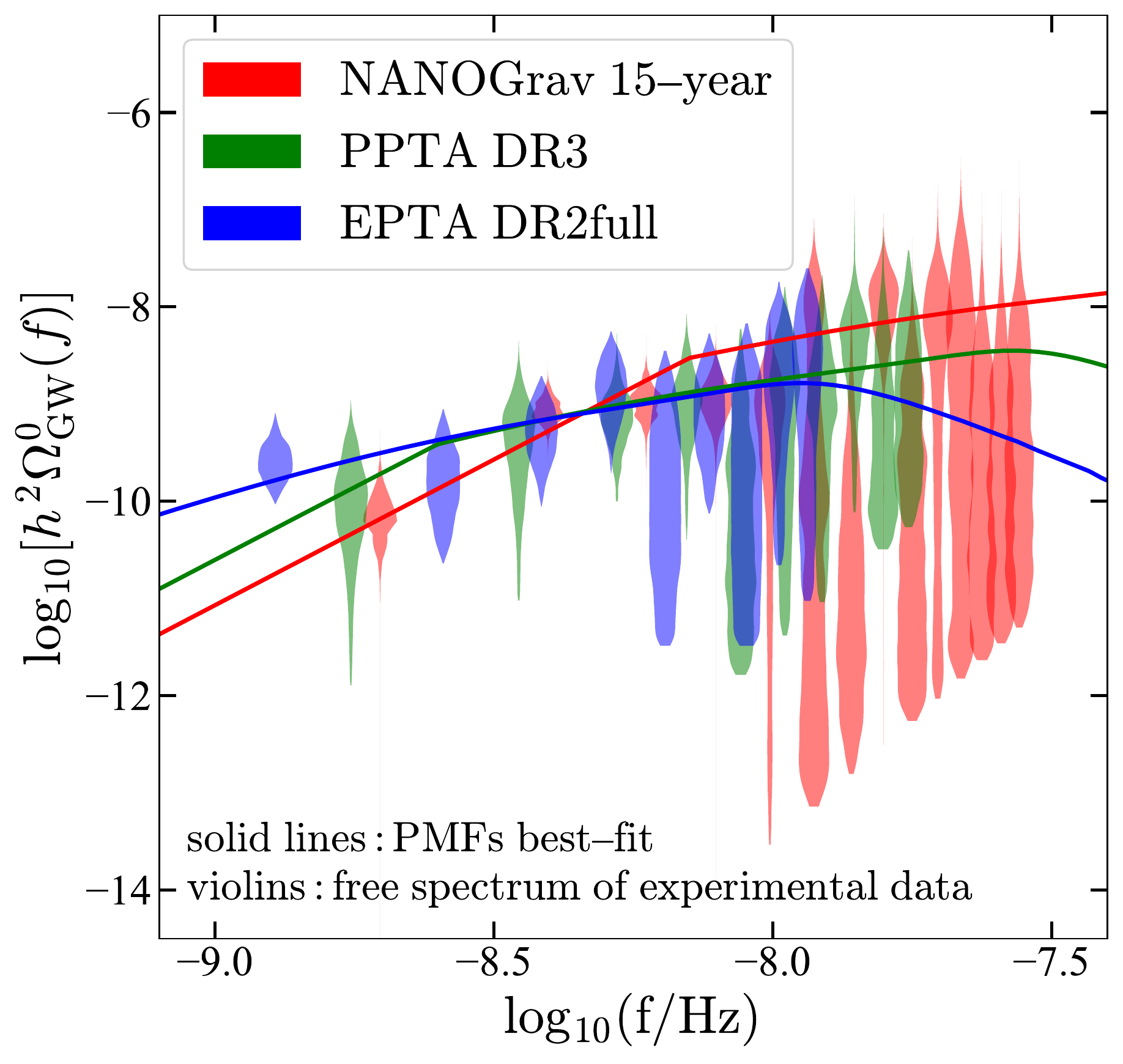}
    \includegraphics[scale = 0.45]{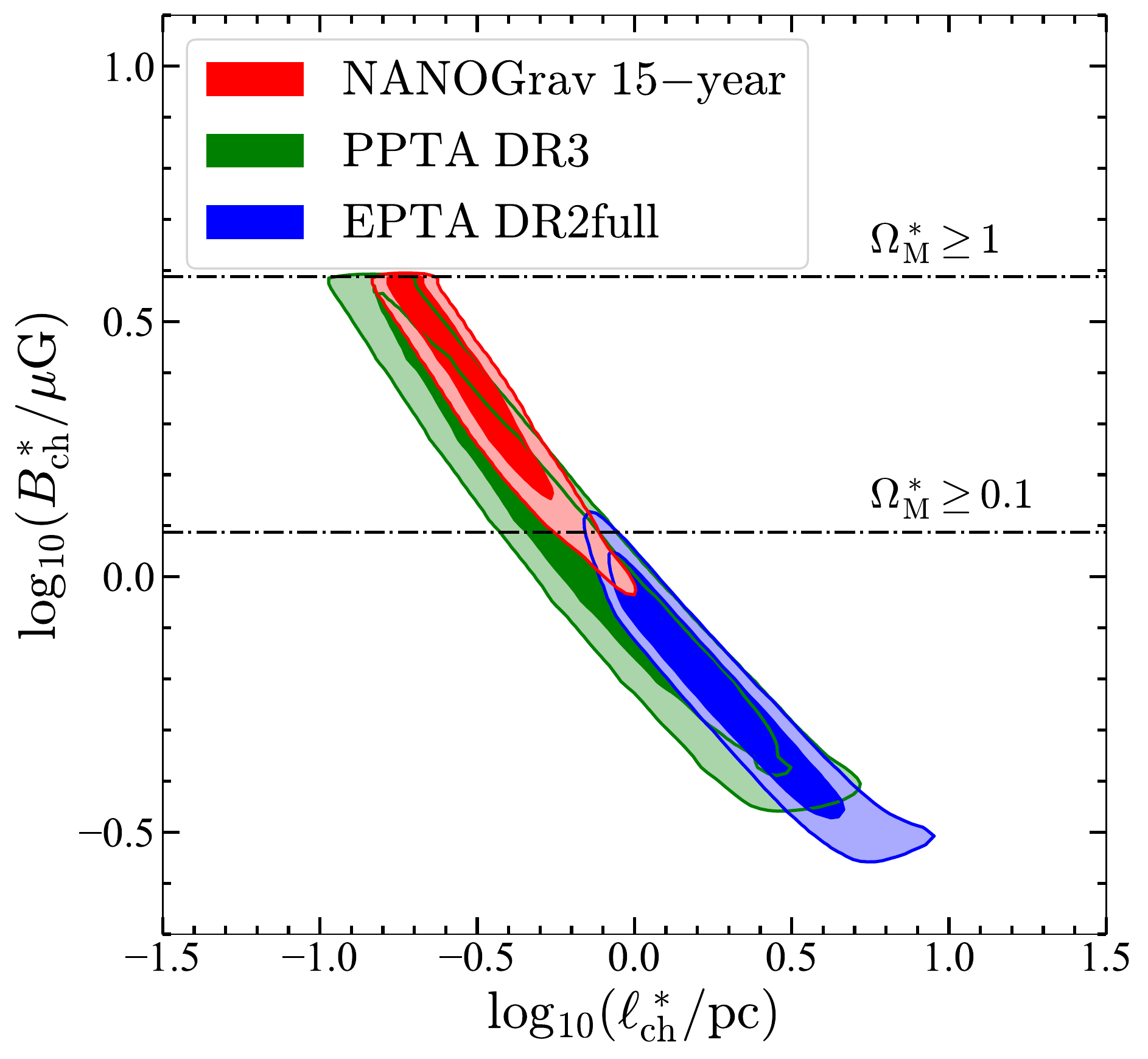}
    \caption{{\bf Left:} the free-spectrum analysis for the NANOGrav 15-year dataset (red), PPTA DR3 (green), and EPTA DR2full (blue) and the theoretical SGWB spectra with the best-fit values of PMFs parameters. {\bf Right:} the favored $68\%$ (inner) and $95\%$ (outer) parameter regions 
    projected on the ($\log_{10}\ell_{\rm ch}^*$, $\log_{10}B_{\rm ch}^*$) plane. 
    Those contours are obtained by  
    %obtained through 
    fitting the timing residual signals of corresponding PTA data. 
    The two dashed lines are the upper limits on the total normalized magnetic energy density from the big bang nucleosynthesis (BBN) constraints ($\Omega_{\rm M}^* = 0.1$ ~\footnote{T. Kahniashvili et al. have revisited the BBN constraints on PMFs. They concluded that when considering the MHD turbulent decay process from the time of the generation of PMFs to BBN, $\Omega_{\rm M}^*\ge 0.1$ is also allowed.~\cite{Kahniashvili:2021gym}};\cite{Grasso:1996kk,Kahniashvili:2009qi}) and the total energy density of the Universe at the initial time ($\Omega_{\rm M}^*=1$). 
    }
    \label{Fig:QCD_Bs_ls}
\end{figure*}

A maximum likelihood estimation method is utilized to fit the parameters of the PMF model, $B_{\rm ch}^*$ and $\ell_{\rm ch}^*$, by means of comparing the theoretical SGWB to the common-noise free spectrum derived from PPTA DR3, EPTA DR2full, and NANOGrav 15-year
data \cite{NANOGrav:2023gor,Antoniadis:2023ott,Reardon:2023gzh}. The violin plots in the left panel in Fig.~\ref{Fig:QCD_Bs_ls} depict the free spectra with Helling-Downs cross-correlations, and the solid lines are the analytical SGWB spectra incorporating the best-fit values of $B_{\rm ch}^*$ and  $\ell_{\rm ch}^*$.
The red, green, and blue contours in the right panel in Fig.~\ref{Fig:QCD_Bs_ls} correspond to the obtained 68\% and 95\% confidence level parameter regions of $B_{\rm ch}^*$ and $l_{\rm ch}^*$ using the NANOGrav 15-year data, PPTA DR3, and EPTA DR2full, respectively. 
As shown in this panel, the favored parameters are $B_{\rm ch}^* \sim \mathcal{O}(1)~\rm{\mu G} $ and $l_{\rm ch}^* \sim \mathcal{O}(1)~\rm{pc}$. 
We would note that the three contours are located at slightly different regions because the amplitudes of the free spectra measured by those three experiments are different.

%%%%%%%%%%%%%%%%%%%%%%%%%%%%%%%%%%%%%%%%%%%%%%%%%%%%%%%%%%%%%%%%%%%%%%%%%%%%%%%%
\section{Baryon Inhomogeneities Induced from PMFs at Recombination}
\label{clumping_b}
%%%%%%%%%%%%%%%%%%%%%%%%%%%%%%%%%%%%%%%%%%%%%%%%%%%%%%%%%%%%%%%%%%%%%%%%%%%%%%%%
% In the previous section, we have determined the PMF parameters at the QCD phase transition time.
% Now we discuss the impacts of such PMFs on recombination history. 
% Firstly,
To study the impacts of PMFs on recombination history, we calculate the characteristic magnetic field strength right before recombination $B_{\rm ch}^{\rm rec}$, by combining the linear relation between $B_{\rm ch}^{\rm rec}$ and $l_{\rm ch}^{\rm rec}$ and the evolution track $B_{\rm ch}\sim\ell_{\rm ch}^{-\alpha}$~\cite{Banerjee:2004df,Subramanian:2015lua,Jedamzik:2010cy,Durrer:2013pga}:
\begin{equation}
    \frac{B^{\rm rec}_{\rm ch}}{\rm nG}= \left[\frac{B_{\rm ch}^*}{\rm nG}\times 
    \left(\frac{\ell_{\rm ch}^*}{0.1~\rm Mpc}\right)^\alpha\right]^{\frac{1}{\alpha+1}},
    \label{Brec_Bs_ls_alpha}
\end{equation}
where $\alpha$ characterizes the cascade process. Theoretically, it is determined by the ideal invariant in the MHD system, i.e., $\alpha = 3/2$ from Saffman flux invariant for compressible fluids, $\alpha = 1/2$ from helicity conservation for helical magnetic fields, $\alpha = 5/4$ from Saffman helicity invariant\cite{Hosking:2020wom,Zhou:2022xhk, Brandenburg:2022rqn}, etc. The ideal invariant in the MHD system with PMFs is unclear. Therefore, we set $\alpha$ as a free parameter in this paper. 

This magnetic field $B_{\rm ch}^{\rm rec}$ can induce the baryon density inhomogeneities at recombination~\cite{Jedamzik:2018itu}. 
We define a baryon clumping factor to represent the density inhomogeneities as
\begin{equation}
    b=\frac{\langle(n_{\rm b} - \langle n_{\rm b} \rangle)^2\rangle}{\langle n_{\rm b} \rangle ^2}. 
\end{equation} 
Following Ref.~\cite{Jedamzik:2018itu}, $B_{\rm ch}^{\rm rec}$ and $b$ can be linked through the $\rm{Alfv\acute{e}n}$ wave speed $c_{\rm A}$, 
namely, $b\simeq \rm{min}[1,(c_{\rm A}/c_{\rm s})^4]$, where $c_{\rm A} = (4.34~{\rm km/s})\times [B_{\rm ch}^{\rm rec}/0.03~\rm{nG}]$ and $c_{\rm s} = 6.33~\rm{km/s}$ at $z=1090$. 
This correlation, however, deviates significantly from the simulation results when $c_{\rm A}/c_{\rm s}$ approaches one~\cite{Jedamzik:2018itu}. 
As a result, we adopt the relation between $c_{\rm A}/c_{\rm s}$ and $b$ from the simulation results given in Ref.~\cite{Jedamzik:2018itu}.

The recombination epoch is dominated by two processes: hydrogen recombination and ionization. 
The recombination rate is proportional to the squared electron density $n^2_{\rm e}$, and the ionization rate is proportional to the neutral hydrogen density $n_{\rm HI}$. 
Incorporating a baryon density fluctuation, induced by $B_{\rm ch}^{\rm rec}$, leads to an inhomogeneous Universe with $\langle n^2_{\rm e}\rangle > \langle n_{\rm e} \rangle^2$. 
This enhances the average recombination rate, leading to earlier recombination and a reduction in the CMB sound horizon $r_{\rm sh}$.
Given a fixed angular sound horizon $\theta_{\rm ls}$, the conformal distance to the CMB $r_{\rm ls}$ decreases simultaneously because of $\theta_{\rm ls} \propto r_{\rm sh}/r_{\rm ls} $.
Such a smaller $r_{\rm ls}$ implies a larger $H_0$ value.

To qualitatively estimate the impacts of $b$ on the recombination process, 
we use a three-zone model~\cite{Jedamzik:2020krr}. 
We adopt a benchmark model M1 from Ref.~\cite{Jedamzik:2020krr} 
by setting the volume fraction of the second zone $f_2=1/3$, density parameters for the first zone $\delta_1=0.1$, and 
density parameters for the second zone $\delta_2=1$, based on the reason that it alleviates $H_0$ tension better than the model M2. 
Instead of treating $b$ as a free parameter as given in Ref.~\cite{Jedamzik:2020krr}, 
we incorporate $b$ as a function of PMF parameters ($B_{\rm ch}^*$, $\ell_{\rm ch}^*$, and $\alpha$) when performing our statistic analysis.

Embedding a modified version of {\fontsize{7.5}{10}\selectfont{CLASS}} 
code~\cite{Rashkovetskyi:2021rwg,Blas:2011rf} 
into {\fontsize{7.5}{10}\selectfont{MontePython}}~\cite{Brinckmann:2018cvx,Audren:2012wb}, a Monte Carlo code for cosmological Bayesian analysis, 
we compute the posterior distributions for $B_{\rm ch}^*$, $\ell_{\rm ch}^*$, $\alpha$ and other cosmological parameters. The experiments used for comparison include Planck 2018 (high TT, TE, EE + low EE, TT + lensing)~\cite{Planck:2018vyg}, BAO~\cite{BOSS:2016wmc,Beutler:2011hx,Ross:2014qpa}, NANOGrav 15-year, PPTA DR3, and EPTA DR2full.  
For addressing the Hubble tension, H3 (SH0ES, MCP, and H0LiCOW) is also included.  
The priors of input parameters are shown in Table~\ref{tab:mcmc_prior}.
\begin{table}[htb!]
    \centering
\begin{tabular}{c|c|c}
\hline
\hline
    Parameter & Prior distribution & Prior range \\
\hline
\hline
\multicolumn{3}{|c|}{\textbf{$\Lambda$CDM Cosmology}}\\
\hline
\hline
    $\Omega_b h^2$ & Flat & [0.02, 0.02] \\
    $\Omega_{\rm cdm} h^2$ & Flat & [0.11, 0.13] \\
    $100 \cdot \theta_s$ & Flat & [1.04, 1.04] \\
    $\ln\left(A_s\times 10^{10}\right)$ & Flat & [2.96, 3.14] \\
    $n_s$ & Flat & [0.94, 0.99] \\
    $\tau_{\rm reio}$ & Flat & [0.01, 0.70] \\
\hline
\hline
\multicolumn{3}{|c|}{\textbf{PMFs}}\\
\hline
\hline
    $B^*_{\rm ch}$ & Log & [$10^{-2}$, $10^{0.59}$] \\
    $\ell^*_{\rm ch}$ & Log & [$10^{-2}$, $10^{1.50}$] \\
    $\alpha$ & Flat & [0.60, 3.00]\\
\hline
\hline
\end{tabular}
    \caption{All the input parameters used in our scan. 
    Two types of parameters are grouped as cosmological and PMF parameters.}
\label{tab:mcmc_prior}
\end{table}

\begin{figure*}[htb!]
  \centering
  \includegraphics[scale = 0.45]{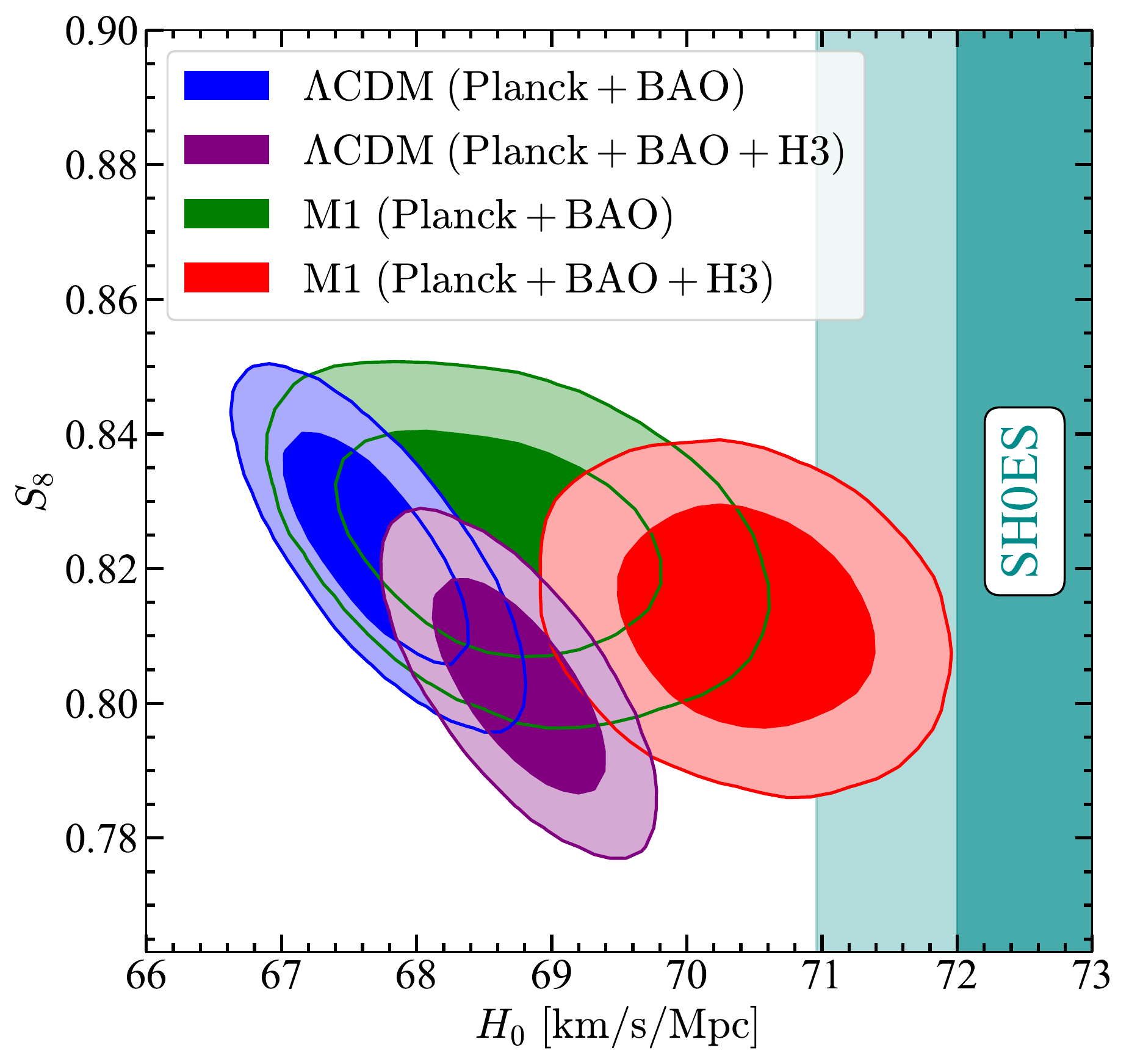}
  \includegraphics[scale = 0.45]{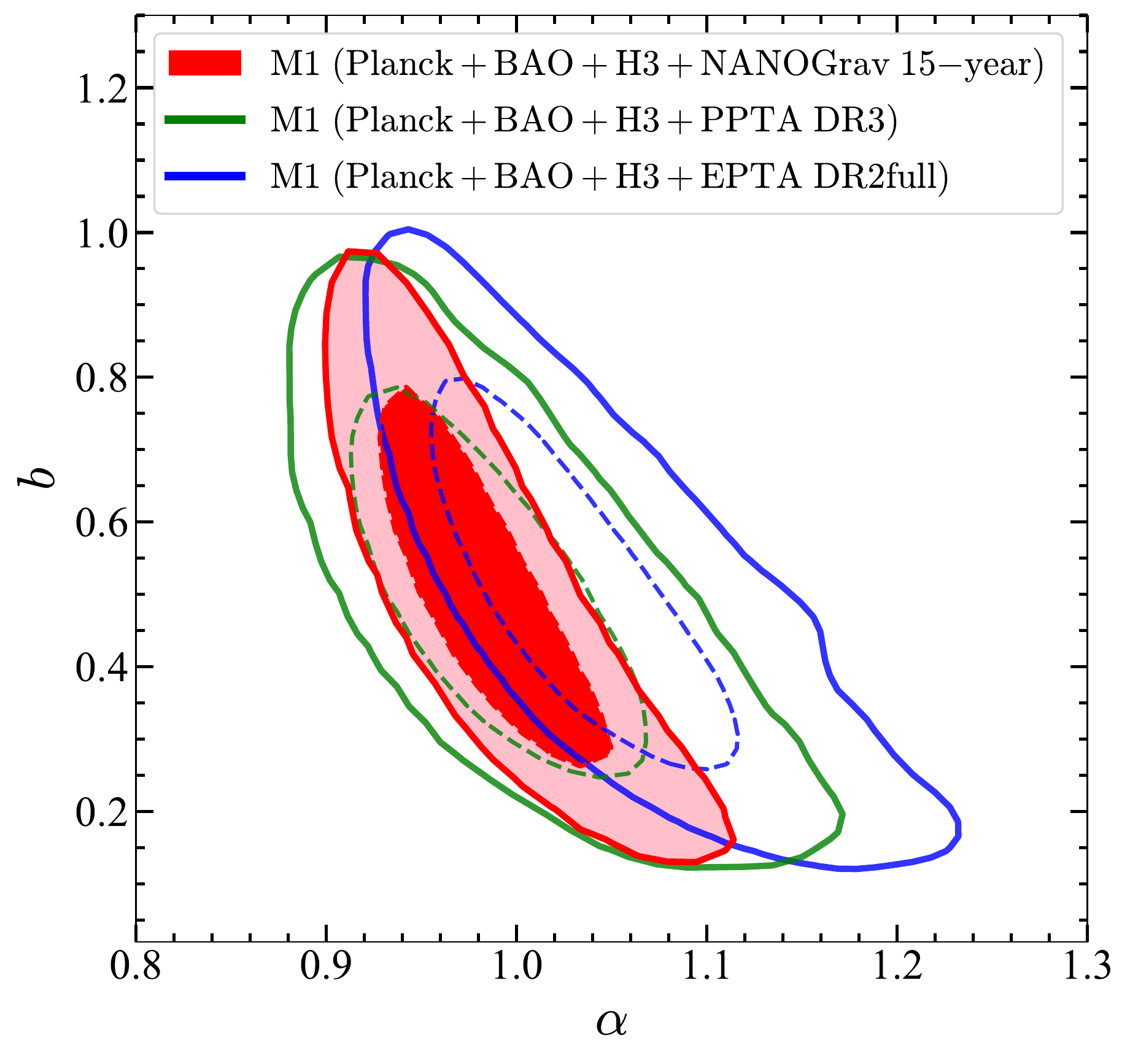}
  \caption{The marginalized $68\%$ (inner contour) and $95\%$ (outer contour) credible regions of $H_0-S_8$ (left panel) and $\alpha-b$ (right panel). The dark cyan region in the left panel shows the $68\%$ and $95\%$ credible regions of $H_0$ from SH0ES. }
  \label{Fig:H0_S8}
\end{figure*}
Figure~\ref{Fig:H0_S8} shows the projected two-dimensional posterior distributions onto the ($H_0$, $S_8$) plane (left panel) and 
($\alpha$, $b$) plane (right panel). We find that the M1 model can increase the $H_0$ value to $70.4\pm0.6$ and reduce the $S_8$ value to $0.813\pm0.01$ for  
alleviating $H_0$ and $S_8$ tension. 
Even without adding H3 likelihood, the M1 model (green contour in Fig.~\ref{Fig:H0_S8}) can also elevate the value of $H_{0}$ in comparison of the $\Lambda \rm{CDM}$ model (blue contour in Fig.~\ref{Fig:H0_S8}).
The right panel in Fig.~\ref{Fig:H0_S8} shows a strong correlation between $b$ and $\alpha$. 
The index $\alpha$ dominates the evolution of the magnetic field, and its impact on the clumping is larger than that of $B_{\rm ch}^*$ and $\ell_{\rm ch}^*$. Therefore, different PTA likelihood leads to little discrepancy between the $\alpha-b$ contours. The M1 model gives $b=0.51^{+0.17}_{-0.18}$ for three PTA likelihoods, but $\alpha=0.99^{+0.03}_{-0.05}$ for including NANOGrav likelihood, $\alpha=1.0^{+0.03}_{-0.05}$ for PPTA, and $\alpha=1.0^{+0.04}_{-0.07}$ for EPTA.

%%%%%%%%%%%%%%%%%%%%%%%%%%%%%%%%%%%%%%%%%%%%%%%%%%%%%%%%%%%%%%%%%%%%%%%
\section{\textbf{Conclusion and Discussion}}
% {\it Conclusion and Discussion.}
\label{c & d}
%%%%%%%%%%%%%%%%%%%%%%%%%%%%%%%%%%%%%%%%%%%%%%%%%%%%%%%%%%%%%%%%%%%%%%%
In this work, we use the NANOGrav~15-year, EPTA~DR2full, and PPTA~DR3 data to constrain the characteristic strength and scale of the PMFs generated in the early Universe, $B_{\rm ch}^* \sim \mathcal{O}(1)~\rm{\mu G}$ and $\ell_{\rm ch}^* \sim \mathcal{O}(1)~\rm{pc}$, assuming that the SGWB can be produced by PMF-induced MHD turbulence. 
In addition, we employ the model parameter $B^*_{\rm ch}$, $\ell^*_{\rm ch}$, and the evolution parameter $\alpha$ instead of clumping factor $b$ for Monte Carlo scan. For likelihoods, we have: Planck 2018 (high TT, TE, EE + low EE, TT + lensing), BAO, H3, NANOGrav 15-year, PPTA DR3, and EPTA DR2full. 
We find that the M1 model prefers a clumping factor $b\sim 0.5$, $\alpha\sim 1$, and $B_{\rm ch}^{\rm rec}\sim 0.1{\rm \ nG}$, 
which helps to alleviate the $H_0$ and $S_8$ tension. Our study reports the first data-driven interval for $\alpha$, which holds significance for researchers investigating theoretical index values.
\begin{figure}[htb!]
  \centering
  \includegraphics[scale = 0.45]{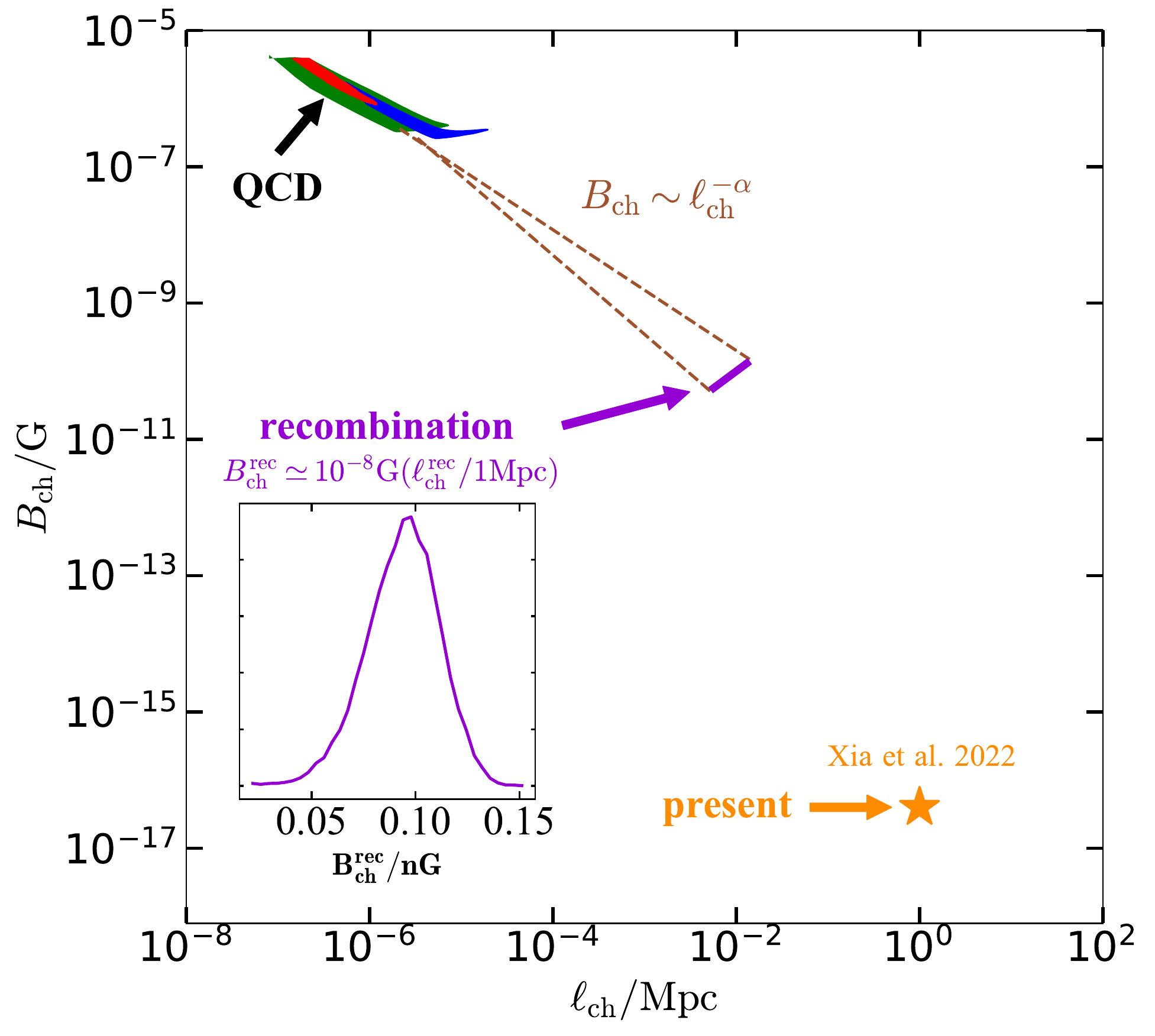}
  \caption{The comoving frame strength and the characteristic length of the magnetic fields inferred in our modeling. 
  The red, green, and blue regions correspond to NANOGrav, PPTA, and EPTA results, respectively, during the QCD phase transition period. 
  The purple region denotes the recombination epoch. 
  The brown dashed line connects the two cosmic times via $B_{\rm ch} \sim \ell_{\rm ch}^{-\alpha}$. 
  The inset plot shows the probability density distribution of $B_{\rm ch}^{\rm rec}$. 
  The present value of $B\sim 4\times 10^{-17}$ G, assuming a coherence length of $\sim 1$ Mpc, is inferred from the gamma-ray observations of GRB 221009A~\cite{Xia:2022uua}. }
  \label{Fig:QCD_rec_now}
\end{figure}

We comment that the recent observations of the high-energy afterglow of GRB 221009A~\cite{Xia:2022uua} by Fermi-LAT may indicate that there was a delayed cascade emission from very-high-energy photons in the background radiation field, which gives a measurement of the intergalactic magnetic fields at a characteristic scale around $\mathcal O(1)$~Mpc with field strength $B_0 \sim \mathcal O(10^{-16})$~G. 
In Fig~\ref{Fig:QCD_rec_now}, we present a summary plot of the magnetic fields at three periods: the QCD phase transition, recombination, and the present epoch. 
This present value is about 5 orders of magnitude lower than that needed to alleviate the Hubble tension at the recombination time, suggesting the presence of a very efficient magnetic energy dissipation process.
We will leave this topic for future studies.

\section*{Acknowledgements}
We thank Ligong Bian, Karsten Jedamzik, Alberto Roper Pol, Ziqing Xia, and Hong-Zhe Zhou for the useful discussion. This work is supported by the National Natural Science Foundation of China (No. 11921003), and the Chinese Academy of Sciences.

\bibliographystyle{JHEP}
\bibliography{bibtex.bib}
\end{document}